\documentclass[copyright]{eptcs}
\providecommand{\event}{F-IDE 2015} 
\usepackage{paralist}
\usepackage{graphicx}
\usepackage{acronym}
\usepackage{xspace}
\usepackage{enumitem}
\usepackage{epstopdf}
\usepackage{cleveref}

\acrodef{vdm}[VDM]{Vienna Development Method}
\acrodef{cml}[CML]{COMPASS Modelling Language}
\acrodef{csp}[CSP]{Communicating Sequential Processes}
\acrodef{utp}[UTP]{Unifying Theories of Programming}
\acrodef{cps}[CPS]{Cyber Physical System}

\acrodef{fm}[FM]{Formal Methods}

\acrodef{ide}[IDE]{Integrated Development Environment}
\acrodef{rcp}[RCP]{Rich Client Platform}
\acrodef{ast}[AST]{Abstract Syntax Tree}
\acrodef{ir}[IR]{Intermediate Representation}
\acrodef{dltk}[DLTK]{Dynamic Languages Toolkit}
\acrodef{dsl}[DSL]{Domain-Specific Language}
\acrodef{ui}[UI]{User Interface}
\acrodef{ct}[CT]{Combinatorial Testing}

\def\astcreator{\textit{AstCreator}\xspace}
\def\xtext{\textit{Xtext}\xspace}
\def\compass{COMPASS\xspace}

\begin{document}

\title{Towards Enabling Overture as a Platform for Formal Notation IDEs}
\def\titlerunning{The Overture Platform IDE}

\author{Lu\'{i}s Diogo Couto \qquad Peter Gorm Larsen \qquad Miran Hasanagi\'{c} \qquad Georgios Kanakis\\
Kenneth Lausdahl \qquad Peter W. V. Tran-J\o{}rgensen
\institute{Department of Engineering, Aarhus University,\\
Finlandsgade 22, 8200 Aarhus N, Denmark}
\email{\{ldc,pgl,miran.hasanagic,gkanos,lausdahl,pvj\}@eng.au.dk}
}
\def\authorrunning{Lu\'{i}s Diogo Couto et. al.}
\maketitle

\begin{abstract}
Formal Methods tools will never have as many users as tools for popular
programming languages and so the effort spent on constructing Integrated
Development Environments (IDEs) will be orders of magnitudes lower than that of
programming languages such as Java. This means newcomers to formal methods do
not get the same user experience as with their favourite programming IDE. In
order to improve this situation it is essential that efforts are combined so it
is possible to reuse common features and thus not start from scratch every
time.  This paper presents the Overture platform where such a reuse philosophy
is present. We give an overview of the platform itself as well as the
extensibility principles that enable much of the reuse. The paper also contains
several examples platform extensions, both in the form of new
features and a new IDE supporting a new language.
\end{abstract}

\section{Introduction}
\label{sec:intro}

The \ac{vdm} is one of the most mature \ac{fm} ~\cite{Jones99,Fitzgerald&09}.
The method focuses on the development and
analysis of a system model expressed in a formal language. The formality of the language enables developers to use a wide range of analytic techniques, from
testing to mathematical proof, to verify the consistency of a model and its
correctness with respect to an existing statement of requirements.  The \ac{vdm}
modelling language has been gradually extended over time. Its most basic
form~(VDM-SL), standardised by ISO~\cite{ISOVDM96short} supports the modelling
of the functionality of sequential systems. Extensions support object-oriented
modelling and concurrency (VDM++)~\cite{Fitzgerald&05}, real-time
computations~\cite{Mukherjee&00} and distributed
systems (VDM-RT)~\cite{Verhoef&06b,Verhoef09}. All these dialects of \ac{vdm} are supported
by the Overture platform~\cite{Larsen&10adoi}.\footnote{See
    \url{http://overturetool.org}.}

The mission of the Overture open-source project is twofold:

\begin{itemize} \item To provide an industrial-strength tool that supports the
            use of precise abstract models in any \ac{vdm} dialect for software
    development.  \item To foster an environment that allows researchers and
        other interested parties to experiment with modifications and
        extensions to the tool and the different \ac{vdm} dialects.  
\end{itemize}

\begin{figure}[!Hpt]
\center \includegraphics[width=9cm]{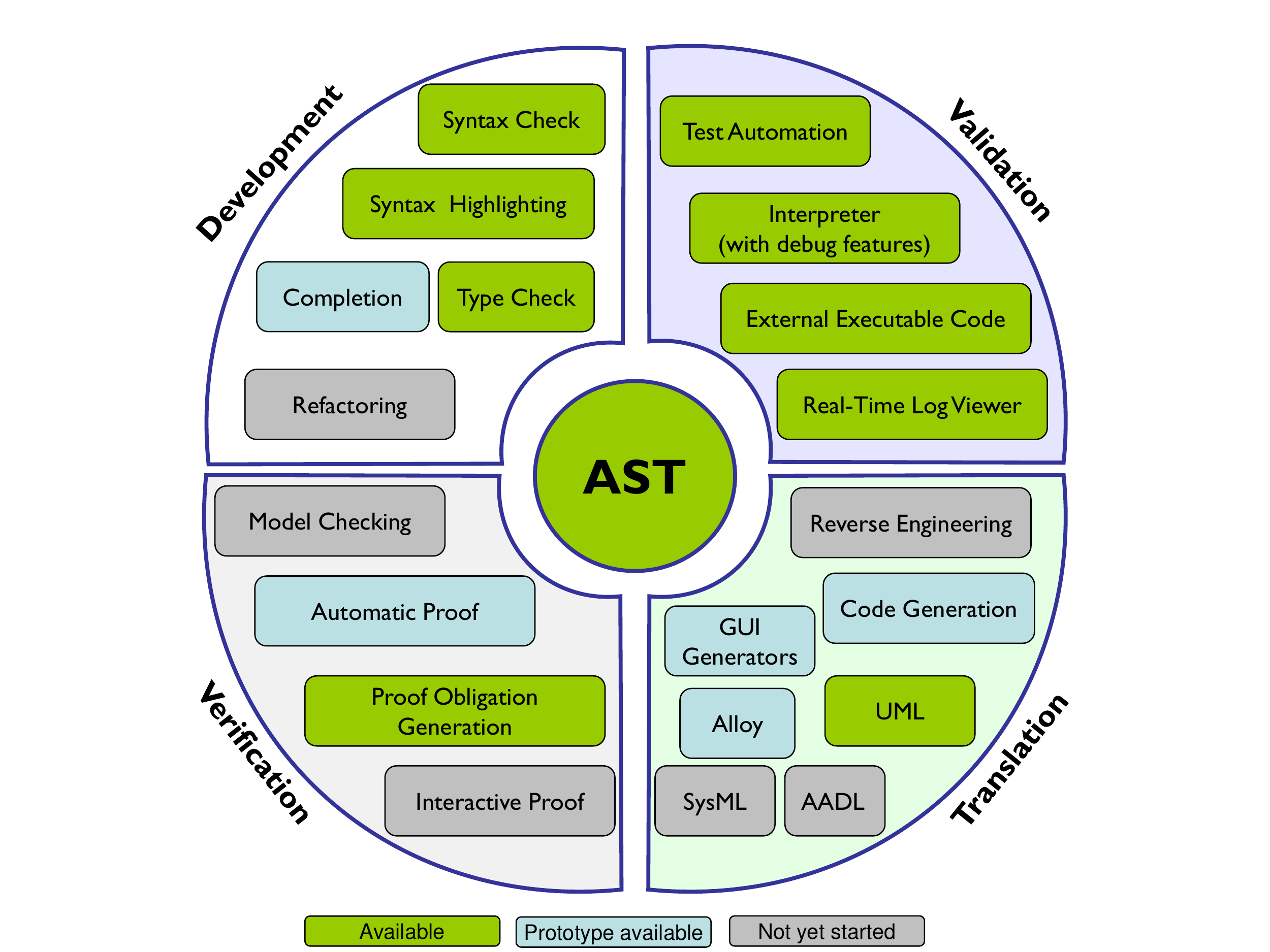}
\caption{Overture Tool Components}
\label{fig:overturetool}
\end{figure}

As is the case with other \ac{fm} tools, the Overture \ac{ide} consists of a
common \ac{ast} representing the model and various plug-ins providing the
different kinds of analysis available in \ac{vdm} as shown in
\cref{fig:overturetool}. The broad variety of analysis possible is common in
many formal methods. In such cases, it is important to ensure that all analyses
are implemented in a consistent way to facilitate maintenance. Such consistency
would also aid in the development and integration of new functional extensions.

In addition to the above, this platform-based architecture allows for the reuse
of common features across all extensions. This reuse can be taken further by
supporting language extensions that would allow other formal notations to reuse the same platform. Recently, Overture has been re-factored to enable such
a reuse \cite{Coleman&12b,Couto&15a}. The main contribution reported in this
paper is the Overture platform itself and its extensibility principles which
are described in sections~\ref{sec:Overture} and~\ref{sec:ExtPrinciples}. An
extensible platform facilitates the development of new features for an \ac{ide}
and in \cref{sec:plugins} we demonstrate how several features of the
Overture \ac{ide} have been developed on top of the platform. Furthermore,
in \cref{sec:Crescendo} we demonstrate how the platform has been
integrated with an external tool.

The extensibility principles of the Overture platform also affect the notation
itself. The platform is capable of supporting a base language (\ac{vdm} in the
case of Overture) as well as multiple notation extensions. This allows for the
development of \acp{ide} for new notations with heavy reuse of common features.
\Cref{sec:Symphony} describes one such \ac{ide}, which also includes
integration with several external tools.

Open issues remain in the platform, most notably in terms of integration with
external tools. \Cref{sec:Conclusion} lays out future work for addressing some
of these issues and also summarises the paper.  It is our hope that this paper
demonstrates the advantages of platform-based \ac{ide} development and that it
can be beneficial for multiple \ac{fm} tool builders to share a common
platform.

Other examples of \ac{fm} platforms with comparable functionalities include
the Asmeta tool set for ASM \cite{ASM,AsmWS15}, the Rodin platform for Event-B
\cite{EventB,Rodin,EventBWs15} or TLAToolbox for TLA$^+$ \cite{TLAplus,TlaWS15}.  The extension
philosophies of the software tools differ as do the actual extensions that are
available.  A detailed comparison is beyond the scope of this article, but more information about these platforms and the modelling languages they support can be found by following the references provided.
The general philosophy of reuse has also been employed effectively for
theorem provers \cite{Paulson10}.

\section{The Overture Platform}
\label{sec:Overture}

\subsection{Overview}

\begin{figure}[!Hpt]
    \begin{center}
        \includegraphics[width=.8\textwidth]{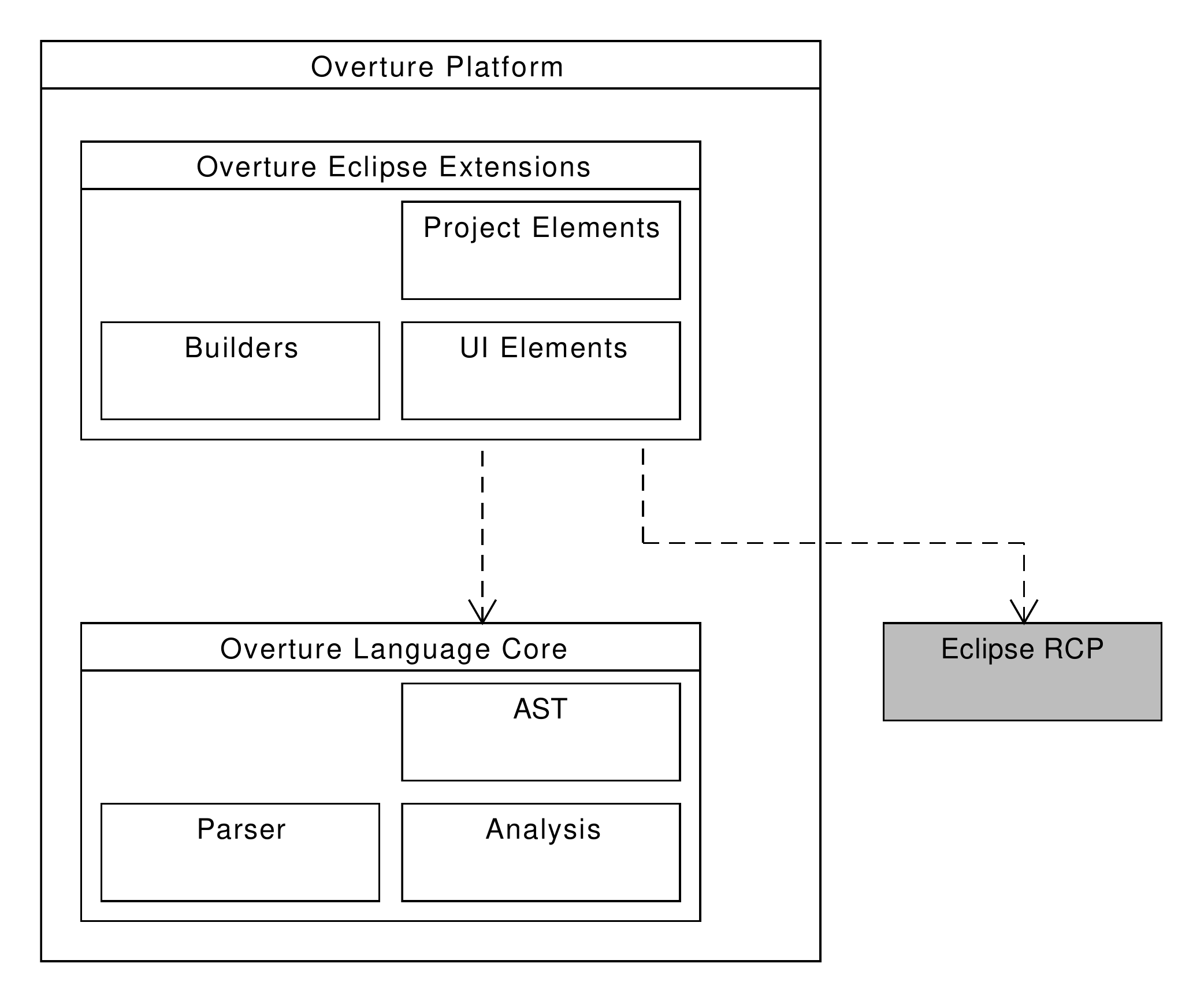}
    \end{center}
    \caption{The Overture Platform.}
    \label{fig:platform-archi}
\end{figure}

The Overture platform supports the development of \ac{fm} \acp{ide}. It
was originally developed to support the development of the Overture \ac{ide} for
\ac{vdm} but has since evolved into a more general platform. It is
comprised of two parts: the \emph{Overture Language Core} and the
\emph{Overture Eclipse Extensions}, as shown in
\cref{fig:platform-archi}.

\subsection{Overture Language Core}
The language core encapsulates and handles any language and notation-related
concerns, including parsing, representation and analysis, in order to facilitate decoupling
between the core language and \ac{ui} implementations. In addition to the general benefits of 
separation of concerns, the language core also opens the possibility of migrating the \ac{ide}
implementation to another \ac{ui} technology as well as providing the base tool
functionalities for command line access, batch processing or as an external tool to
be accessed by others.

The language core consists of an extensible \ac{ast} that is automatically
generated by the \astcreator 
tool\footnote{See \url{http://github.com/overturetool/astcreator}},
as well as a parser for constructing the \ac{ast} from model sources. In
addition, \astcreator also generates machinery for traversing and processing
trees in a consistent way in the form of a visitor
framework~\cite{DESIGNPAT95}. Any kind of analysis of the \ac{ast} such as type
checking or interpretation should be implemented using the visitor framework.

One of the key features of the language core is its extensibility mechanism
which allows language extensions or new languages to be implemented in the
Overture platform while reusing as much existing code as possible. This
mechanism is described in further detail in \cref{sec:ExtPrinciples}

\subsection{Overture Eclipse Extensions}

The Overture platform also consists of a set of extensions to the Eclipse
\ac{rcp} that are used to help build the \ac{ui} components of the \ac{ide}.
The Eclipse \ac{rcp} is a generic framework for building rich client
applications using the Eclipse OSGi plug-in model and \ac{ui} toolkits. It is
is powerful and generic but comes with a cost: significant amounts of
boilerplate source code and configuration files must be written in order to
prepare it to build an \ac{ide}.

The Overture Eclipse extensions automate some of the configuration and
preparation work by providing the aforementioned boilerplate code targeting
\ac{fm} notations.  The extensions provides an extensible application framework
on top of the \ac{rcp}. It significantly reduces the amount of code that needs
to be written in order to contribute an extension to the \ac{ide}. To put it
another way, the \ac{rcp} API is very wide and the Overture Eclipse Extensions
summarise a portion of it, thus giving developers faster access to the
functionality at the cost of some flexibility. However, the Overture extensions
are fully interoperable with the \ac{rcp} so any other extension that requires
direct access to the \ac{rcp} can still be used. 

There are other frameworks similar to the Overture extensions in the Eclipse
project, such as the \ac{dltk}~\cite{website:dltk} and
\xtext~\cite{website:xtest}. \ac{dltk} is designed to support the
implementation of \acp{ide} for dynamic programming languages and \xtext is
designed to support the implementation of \acp{ide} for for \acp{dsl} or small
programming languages. Neither framework is particularly suitable for \ac{vdm}
-- \ac{vdm} is similar in notation to a statically typed general-purpose
programming language -- which was the original target language to be supported
by the Overture \ac{ide}.

Broadly speaking, the Overture Eclipse extensions can be divided into three groups:

\begin{itemize}
    \item a set of \textsf{UI elements} for editors, launch configurations, etc. that
        interact directly with the Eclipse \ac{rcp}.
    \item a set of \textsf{project elements} that represent the \ac{fm} model
        and associated concepts such as source units, according to the Eclipse project
        model. Also included are connectors and providers for accessing these various
        entities from within the \ac{ide}.
    \item a set of \textsf{builders} that interact with the language core in order to process
        language sources to construct an internal representation of the model and load
        it into the \textsf{project elements}.
\end{itemize}

Both the \textsf{builders} and the \textsf{project elements} are developed according
to standard Eclipse conventions so that new versions of these packages for other
notations may be contributed.

Currently, the Overture Platform primarily supports the Overture \ac{ide}. The Overture
\ac{ide} is comprised of Eclipse plug-ins that use components implemented with
the language core to perform analysis of the \ac{vdm} \ac{ast} and \ac{ui} components that
wrap the analysis and use the Eclipse extensions to implement the interaction with the
user.

\section{Extension Principles of the Overture Language Core}
\label{sec:ExtPrinciples}

The basic principles of extensibility in the Overture language core are related
to the generation of \acp{ast} from specification files, similar to parser
generators like SableCC~\cite{Gagnon&98}. In addition to generating the classes
representing the tree structure, it is important to generate auxiliary
machinery to allow developers to implement analysis of the \ac{ast} in a
consistent manner.

The main way to construct extensions in the language core is by extending the
\ac{ast}.  Generally speaking, an \ac{ast} is extended by adding
new subtrees that are either entirely new or that contain some existing base
nodes. In addition, the extended tree needs to reuse the existing base node
classes wherever possible.

In addition to extending the tree itself, it is important to also
extend the analysis machinery. Particularly, this extended machinery
needs to be able to analyse trees made up of extension and base
nodes. Furthermore, the extended analysis machinery needs to reuse
the base machinery when processing base nodes -- this is essential for
achieving reuse of functionalities already implemented as base analysis.

Whether speaking of a tree made of only extension nodes or a hybrid tree with
extension and base nodes or even a base tree, the \ac{ast} classes have a
limited ability to enforce the structure of each particular instantiation of
the tree. It is the syntax of the language, as encoded in the parser, that
ultimately controls which trees are admissible.  Along the same lines, it is
the parser that controls which base nodes are reused when constructing hybrid
trees as the extended tree specification can only set an upper limit on this.

\begin{figure}[!ht]
  \begin{center}
      \includegraphics[width=0.8\textwidth]{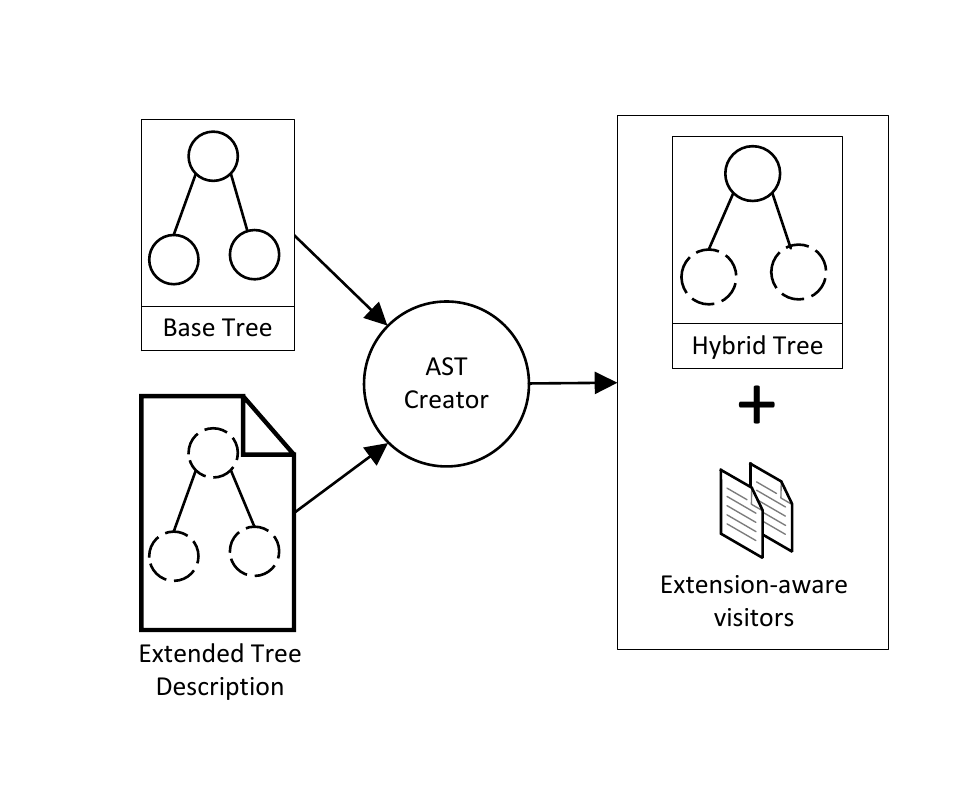}
      \caption{Extending \ac{ast} specifications.}
    \label{fig:ast-ext}
  \end{center}
\end{figure}

The extensibility principles of the Overture language core are primarily
realised through the \astcreator tool. \astcreator provides provides an
automated way of generating trees and auxiliary machinery from specification
files as shown in \cref{fig:ast-ext}. The tool is capable of taking an
existing \ac{ast} as well as an extension specification and generating the
extension nodes and visitor framework upon which to implement
analyses.\footnote{\astcreator is also capable taking a base and extension
specification and producing both sets of classes, though this is done less
frequently.} Both nodes and visitors are aware of the base classes thus
ensuring interoperability with the base trees.

It is also possible to use \astcreator to build a completely new \ac{ast}
supporting a language that is unrelated to \ac{vdm} (see \cref{sub:codegen}
for an example). In this case, a new base tree and visitor framework will be
produced and it will not be possible to reuse existing components of the
language core.  As such, we focus on the case where the new language being
supported is an extension of an existing notation where it is possible to reuse
parts of the base \ac{ast}, and the corresponding analysis. Typically,
constructs like arithmetic or logical expressions or imperative statements can
be reused. This leads to hybrid trees where nodes from the base and extended
trees are blended together.

\begin{figure}[ht]
  \begin{center}
      \includegraphics[width=0.9\textwidth]{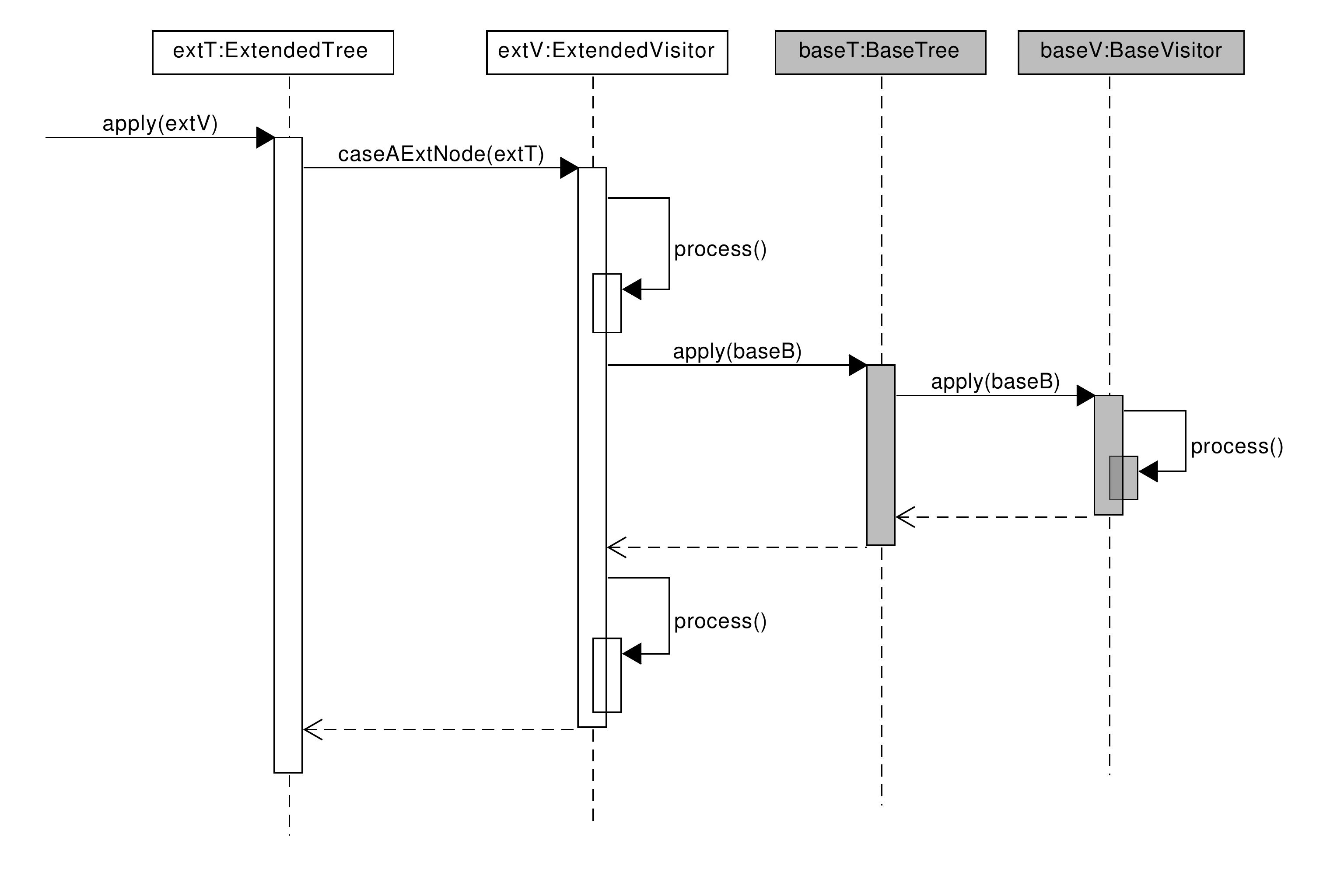}
    \caption{Extended AST and analysis.}
    \label{fig:ext-analysis}
  \end{center}
\end{figure}

The semantics of such a language extension should be implemented as various
\ac{ast} analyses such as type checking or interpretation. Each analysis should
be implemented as an independent component that processes the tree in a
consistent way. The visitor framework that is generated as part of the extension
provides a way to achieve this. Since the visitor framework itself is
extension-aware it enables selective and controlled reuse of existing base
analyses as necessary. The extension-aware visitor is illustrated in
\cref{fig:ext-analysis}. 

An example of these extension principles at work can be seen in the
Symphony IDE, as described in \cref{sec:Symphony}.

\section{Functionality extensions}
\label{sec:plugins}

New functionality can be contributed in the Overture platform either by using
the language core, the Eclipse extensions or a combination thereof. The
language core provides the necessary mechanisms to interact with the \ac{ast}
as well as extending it, whereas the Eclipse extensions provide the means to
expose functionality to the user. In this section, we provide examples of how
both can be used to add new functionality to Overture.

\subsection{The code generation platform}
\label{sub:codegen}

The code generation platform aims to facilitate integration of \ac{vdm} code
generators into Overture with minimum effort~\cite{Jorgensen&14a}. Like many
other Overture components, the code generation platform interacts with the
language core by analysing a type checked \ac{vdm} \ac{ast} in order to generate
code in some target language. Currently, the code generation platform is used
to develop \ac{vdm} code generation support to Java and C++, and in addition,
there is ongoing work on generating Isabelle/HOL syntax~\cite{Couto&15b}.

In order to promote reuse the code generation platform works with an \ac{ir} of
the generated code, which is independent of any particular target language. In
addition, the code generation platform provides mechanisms for rewriting or
\emph{transforming} the \ac{ir} into a semantically preserving form that is
easier for a particular backend to code generate. Furthermore, since
transformations work directly on the \ac{ir} it becomes easier for different
backends to use and contribute new functionality to analyse and modify the
\ac{ir}.

The code generation platform allows new nodes to be added to the \ac{ir} as
well as extending existing nodes with additional fields as enabled by
\astcreator, which is used for the specification of the \ac{ir}. The
Isabelle/HOL code generator exploits this, since mutually recursive functions
must be grouped explicitly in Isabelle/HOL and therefore the code generator
adds function groups to the \ac{ir}. This is done in a supplementary tree
extension file and demonstrates how users of the code generation platform can
extend and change the \ac{ir} as needed. Although most of the work involved in
developing code generation support includes traversing and transforming the
\ac{ir}, thus interacting with the language core, the Eclipse extensions
provide the necessary mechanisms to read preferences and configure the code
generation process.

\subsection{Interpreting implicit specifications using ProB}

In \ac{vdm} functions and operations can be either explicitly or implicitly (using pre and post conditions)
defined. An explicit description defines how the output is obtained from the
input, which enables the description to be evaluated directly in the \ac{vdm}
interpreter~\cite{Lausdahl&11}. Implicit descriptions, on the other hand, only
specify the constraints that must be met but without defining how the output is
obtained. Therefore, attempting to evaluate an implicit description in the
interpreter yields a runtime error. To avoid restricting analysis of implicit
descriptions to static analysis only, Overture has recently integrated the ProB
constraint solver in order to enable evaluation of implicit
descriptions~\cite{Lausdahl&14a}.

Interpretation of implicit descriptions adds an additional step to the model
execution where the pre and post state as well as the constraints imposed by
the implicit description, is converted to ProB syntax to form a formula, which
is submitted to the ProB constraint solver. This formula is constructed as a
string by analysing the type checked \ac{ast}.

If ProB is able to find a solution to the given problem the solution is
converted back to \ac{vdm} format and used throughout subsequent execution of
the model. Intercepting the interpretation of implicit descriptions is
primarily enabled through extension of the language core, and interacting with
ProB via an external Java API.

\subsection{VDMTools integration}

VDMTools \cite{Fitzgerald&08a} is an industrial strength \ac{ide}, maintained
by the SCSK corporation, for analysing models written in \ac{vdm}. Among many
features also supported by Overture, VDMTools provides extensive semantic
checking, execution and Java/C++ code generation of models written in \ac{vdm}.
To facilitate use of different \acp{ide}, Overture provides an option to export
an Overture \ac{vdm} project to a VDMTools compatible format. This plugin is
developed through the Eclipse extensions that are used to convert meta data
from an Overture project to a format compatible with VDMTools.

\subsection{Combinatorial Testing}

\ac{ct} in \ac{vdm} provides automated generation and execution of a large
collection of tests as an extension to the language core
functionality~\cite{Larsen&10cdoi}. The addition of \ac{ct} in Overture has
trigged several changes to the language core components. First, the \ac{ast} was
extended to support the trace nodes. Second, the type checker was updated to
support type checking of both traces and generated tests. Finally, the
interpreter was extended to support trace expansion as well as test execution.

In addition to extending the language core a \ac{ct} view has been added to
Overture as a new Overture Eclipse plugin extension. This plugin serves to
provide a convenient way for users to inspect the test execution results,
filtering large collections of tests in order to obtain a reduced representable
subset of tests, and re-executing tests individually.

\section{Building a Co-Simulation tool with the Overture Platform} 
\label{sec:Crescendo}

The Crescendo tool supports
collaborative modelling and co-simulation of \acp{cps}~\cite{Fitzgerald&14c},
and has been developed by extending the Overture platofrm. This extension enables
co-simulation between co-models, which are composed of a discrete time model
described in the VDM-RT language, and a continues time model described using
differential equations. The extension is composed of a co-simulation engine
that connects an extended version of the \ac{vdm}
interpreter~\cite{Lausdahl&11} from the Overture tool with the simulator in
20-Sim~\cite{20sim}.

The Crescendo tool primarily extends the Overture platform using ordinary
Eclipse extension points for: builders, debug related \ac{ui} and views.
However, it also uses Overture Eclipse extensions for e.g. editors, and
debugging related components.

An extension was also made to the language core by extending the \ac{vdm}
interpreter used for evaluating and debugging with two main features:
\begin{inparaenum}[\itshape a\upshape)] 
\item the ability to only simulate
    until a certain time bound, and 
\item the ability to detect when a shared
    co-simulation variable is accessed.
\end{inparaenum}~This is necessary in order to support co-simulation such that
the two simulators can synchronize their time steps.


\section{Building a new IDE with the Overture Platform}
\label{sec:Symphony}

Thus far, this paper has shown how to contribute extensions to the Overture
\ac{ide} and how to extend existing components to support co-simulation with an
external tool.  These examples consist of extensions that either make very
small or no changes to the \ac{vdm} language, in terms of new syntax, the
semantics thereof or the concepts introduced. This makes the extensions
relatively simple to support in comparison to an extension for a new full-blown
\ac{fm} notation, especially considering the wide variety of formal notations
as well as their associated semantics, paradigms and problem domains.

It is possible to use the Overture platform to build an \ac{ide} for a new
notation that shares nothing with the \ac{vdm} language. However, this means
that the new \ac{ide} will be unable to reuse much of the language core since
the \ac{ast} and associated analyses will be entirely different. On the other
hand, when building an \ac{ide} for a notation that reuses or shares parts of
\ac{vdm}, then the relevant parts of the language core can be reused. The
remainder of this section shows how such reuse was achieved in the construction
of the Symphony tool~\cite{Coleman&12a} in the \compass EU FP7 Project.
Symphony supports the \ac{cml} notation~\cite{Woodcock&12a} that was introduced
in the \compass project and combines \ac{vdm} with \ac{csp}~\cite{Hoare78}. 

The syntax of \ac{cml} differs significantly from that of \ac{vdm}, especially as it
relates to the new constructs inherited from \ac{csp}. As such, it was necessary
to construct a parser to recognize \ac{cml} notation. 
Tools such as ANTLR~\cite{Parr07} greatly aid in parser construction and Symphony has
an ANTLR parser built from scratch that processes \ac{cml} sources to construct
\acp{ast} that are compliant with the Overture language core.

The static analysis of \ac{cml} \acp{ast} (type checking and proof obligation
generation) significantly reuses relevant Overture components~\cite{Couto&13a}.
In the case of proof obligations, reuse led to reduction in lines of code from 2596 to 
978 as well as a reduction in duplicate code from 37.2\% to 3.1\%. In general, any
existing analysis for \ac{vdm} was reused whenever possible. A good example lies
in the processing of \ac{vdm} expressions inside \ac{csp} actions -- also an example
of hybrid tree processing.

The validation of \ac{cml} models could not reuse Overture components
so easily since the paradigms of \ac{cml} notation are different from those of
\ac{vdm}. In particular, \ac{cml} is a process algebra and its models are interpreted
as sequences of events, as opposed to \ac{vdm}'s imperative approach based on
state transformations.

Due to the difference in paradigms between the languages, significant portions
of the Symphony interpreter had to be built from scratch.  However, in spite of
the differences in behaviour, the Symphony interpreter still manages to reuse
the Overture one for evaluating expressions and reused statements. 

For all cases of reuse in Symphony, the same basic principle applies: the
extended analysis processes the hybrid tree and when it encounters a base node,
it submits the node to its counterpart base analysis, with a mechanism in place
for the extension to re-assume control and preventing the base analysis from
hijacking the analysis of the remaining tree.

Finally, it is important to discuss the underlying semantics of the various
analyses as it should be ensured that consistent semantics are in place across
all components of the tool, lest errors be introduced in the overall results
due to gaps between the various semantics. In the \compass project this was
addressed by using the Unifying Theories of
Programming~\cite{Hoare&98} that provides a common framework for the various
semantic models used in the project.  This work eventually led to a
mechanisation of a subset of \ac{cml} in Isabelle~\cite{Foster&15}. 

\begin{figure}[ht]
  \begin{center}
    \includegraphics[width=\textwidth]{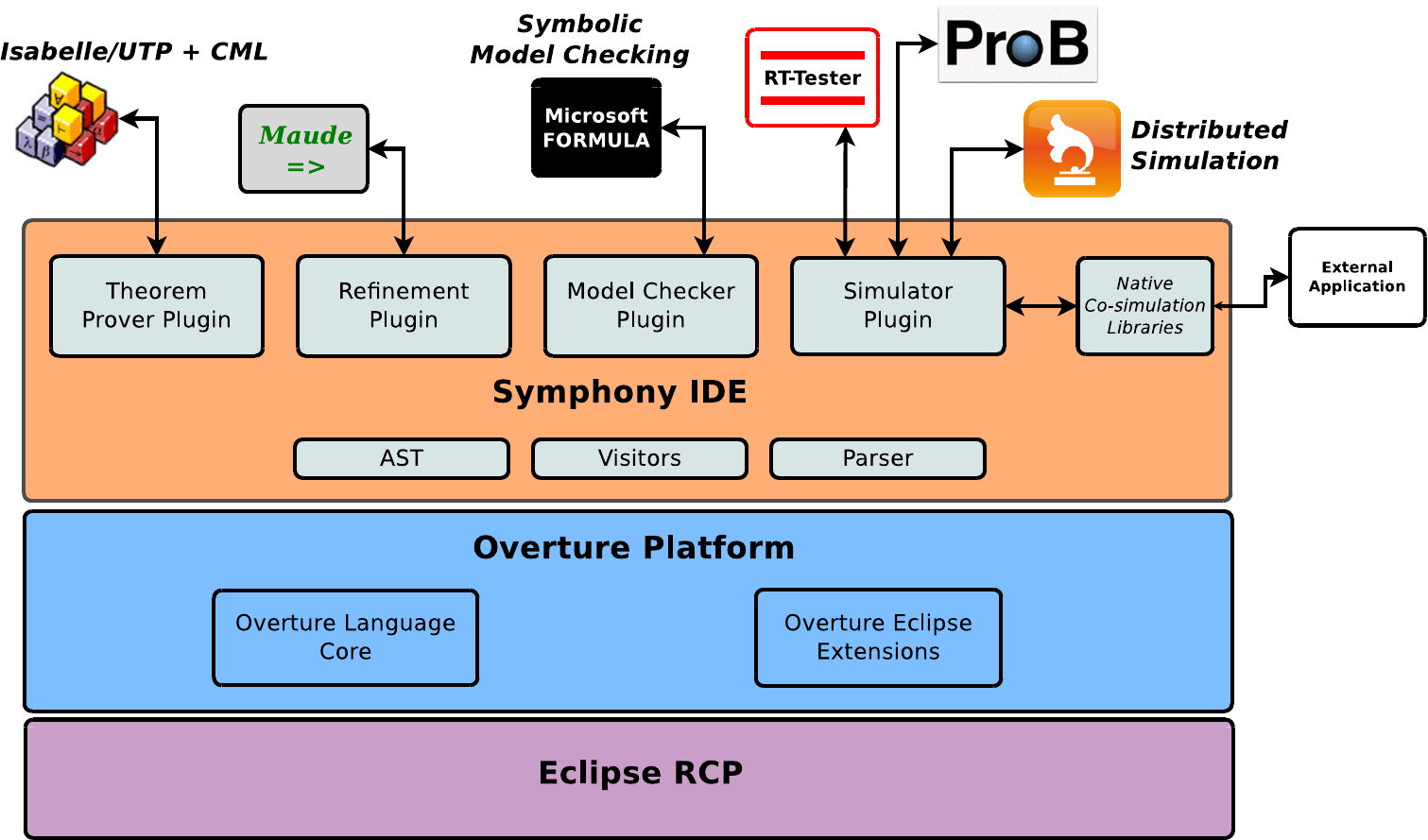}
    \caption{The COMPASS tools}
    \label{fig:compass-archi}
  \end{center}
\end{figure}

Most functionalities of the Symphony \ac{ide} were implemented as Eclipse
plug-ins using a combination of the Overture language core (exported via a counterpart
Symphony core) and the Overture Eclipse extensions (used to build the main \ac{ui} components of
the Symphony \ac{ide}). In addition to its native functionalities, the Symphony
\ac{ide} also uses its various plug-ins to integrate externals tools such as
Maude~\cite{Clavel&99,Clavel&07}, Isabelle~\cite{nipkow02} (via
Isabelle/Eclipse~\cite{website:isabelle-eclipse}), FORMULA~\cite{Jackson&09doi},
RT-Tester~\cite{peleska&11} and ProB\cite{leuschel&03,leuschel&05}.  The most
relevant external tool integrations are shown in \cref{fig:compass-archi}.
Note how this aims at following the principle of reusing existing functionality
rather than re-developing from scratch.

\section{Concluding Remarks and Future Work}\label{sec:Conclusion}

This paper has described the Overture \ac{ide} and its
underlying platform. We have shown the extensibility principles of the
platform and demonstrated how they support multiple functional
extension plug-ins. Furthermore, we have demonstrated how the platform
can support notation extensions and, as such, be used as a basis platform
by other \ac{fm} tool builders. The ability to reuse existing functionality
and build on the work of other teams can help improve the quality of
\ac{fm} tools in general.

Going forward, there are various potential improvements that can be made
to the Overture platform and we discuss a few of them here. The first
improvement is in terms of the \astcreator tool's specification files. At
the moment, \astcreator is only capable of generating the Java code for the
\ac{ast} from a fairly simple tree specification file. This is by design. \astcreator
does not aim to address issues of parsing when tools such as ANTLR already do
an excellent job of it. However, it may be beneficial to integrate \astcreator
with parser generators. Either by deriving an \astcreator specification file from
the parser generator grammar or by creating a stub grammar from the \astcreator
specification.

Another potential improvement lies in making more use of the code generation platform
when integrating external tools. Integration with external tools often consists
of translating the \ac{vdm} syntax into that of the external tool and submitting it to
the tool as is done for example in the ProB integration. These translations are often
implemented manually using the visitor framework. However, by using the code generation
platform, significant gains may be attained in terms of the amount of code that is necessary.
We are currently undertaking work in this direction and early results are very promising
~\cite{Couto&15b}.

The final improvement under consideration is also related to external tool
integration, but is a somewhat open-ended question at the moment. Integration of
external tools is currently done on a case-by-case basis. Each external tool is
integrated in its own way with entirely handwritten code. While the syntax
translation issue may be addressed, the invocation of the external tool, passing
of data to it and collection of results is completely non-standardized. This is
mostly a consequence of all external tools having different ways of accessing
them. However, at a high-level, most external tool interactions can be reduced
to a general case such as external command invocation, protocol-based
communication or API access. It would be beneficial to have mechanisms in the
platform to help deal with each of the general cases. Another alternative
would be a methodological approach where guidelines are produced to help
developers implement each kind of integration in a consistent manner.

\section*{Acknowledgments}

The authors wish to thank Stefan Hallerstede for valuable feedback. Partial
funding for the work reported here was provided by the COMPASS project (Grant
Agreement 287829) as well as the INTO-CPS project (Grant Agreement 644047).

\bibliographystyle{eptcs}
\bibliography{custom,../../../../papers/bib/dan}

\begin{thebibliography}{10}
\providecommand{\bibitemdeclare}[2]{}
\providecommand{\surnamestart}{}
\providecommand{\surnameend}{}
\providecommand{\urlprefix}{Available at }
\providecommand{\url}[1]{\texttt{#1}}
\providecommand{\href}[2]{\texttt{#2}}
\providecommand{\urlalt}[2]{\href{#1}{#2}}
\providecommand{\doi}[1]{doi:\urlalt{http://dx.doi.org/#1}{#1}}
\providecommand{\bibinfo}[2]{#2}

\bibitemdeclare{book}{EventB}
\bibitem{EventB}
\bibinfo{author}{Jean-Raymond \surnamestart Abrial\surnameend}
  (\bibinfo{year}{2010}): \emph{\bibinfo{title}{Modeling in Event-{B} - System
  and Software Engineering}}.
\newblock \bibinfo{publisher}{Cambridge University Press},
  \doi{10.1017/CBO9781139195881}.
\newblock
  \urlprefix\url{http://www.cambridge.org/uk/catalogue/catalogue.asp?isbn=9780521895569}.

\bibitemdeclare{article}{Rodin}
\bibitem{Rodin}
\bibinfo{author}{Jean-Raymond \surnamestart Abrial\surnameend},
  \bibinfo{author}{Michael~J. \surnamestart Butler\surnameend},
  \bibinfo{author}{Stefan \surnamestart Hallerstede\surnameend},
  \bibinfo{author}{Thai~Son \surnamestart Hoang\surnameend},
  \bibinfo{author}{Farhad \surnamestart Mehta\surnameend} \&
  \bibinfo{author}{Laurent \surnamestart Voisin\surnameend}
  (\bibinfo{year}{2010}): \emph{\bibinfo{title}{Rodin: an open toolset for
  modelling and reasoning in Event-{B}}}.
\newblock {\sl \bibinfo{journal}{STTT}}
  \bibinfo{volume}{12}(\bibinfo{number}{6}), pp. \bibinfo{pages}{447--466},
  \doi{10.1007/s10009-010-0145-y}.

\bibitemdeclare{misc}{AsmWS15}
\bibitem{AsmWS15}
 (\bibinfo{year}{2015}): \emph{\bibinfo{title}{{The Asmeta tool set for ASM}}}.
\newblock \bibinfo{howpublished}{\url{http://asmeta.sourceforge.net}}.

\bibitemdeclare{book}{ASM}
\bibitem{ASM}
\bibinfo{author}{Egon \surnamestart B{\"o}rger\surnameend} \&
  \bibinfo{author}{Robert~F. \surnamestart St{\"a}rk\surnameend}
  (\bibinfo{year}{2003}): \emph{\bibinfo{title}{Abstract State Machines. {A}
  Method for High-Level System Design and Analysis}}.
\newblock \bibinfo{publisher}{Springer}, \doi{10.1007/978-3-642-18216-7}.

\bibitemdeclare{inproceedings}{Clavel&99}
\bibitem{Clavel&99}
\bibinfo{author}{M.~\surnamestart Clavel\surnameend},
  \bibinfo{author}{F.~\surnamestart Durn\surnameend},
  \bibinfo{author}{S.~\surnamestart Eker\surnameend},
  \bibinfo{author}{P.~\surnamestart Lincoln\surnameend},
  \bibinfo{author}{N.~\surnamestart Marti-Oliet\surnameend},
  \bibinfo{author}{J.~\surnamestart Meseguer\surnameend} \&
  \bibinfo{author}{J.~F. \surnamestart Quesada\surnameend}
  (\bibinfo{year}{1999}): \emph{\bibinfo{title}{The Maude System}}.
\newblock In: {\sl \bibinfo{booktitle}{Rewriting Techniques and Applications}},
  \bibinfo{publisher}{Springer, LNCS1631}, \doi{10.1007/3-540-48685-2\_18}.

\bibitemdeclare{book}{Clavel&07}
\bibitem{Clavel&07}
\bibinfo{editor}{Manuel \surnamestart Clavel\surnameend},
  \bibinfo{editor}{Francisco \surnamestart Dur{\'a}n\surnameend},
  \bibinfo{editor}{Steven \surnamestart Eker\surnameend},
  \bibinfo{editor}{Patrick \surnamestart Lincoln\surnameend},
  \bibinfo{editor}{Narciso \surnamestart Mart\'{\i}-Oliet\surnameend},
  \bibinfo{editor}{Jos{\'e} \surnamestart Meseguer\surnameend} \&
  \bibinfo{editor}{Carolyn~L. \surnamestart Talcott\surnameend}, editors
  (\bibinfo{year}{2007}): \emph{\bibinfo{title}{All About Maude - A
  High-Performance Logical Framework, How to Specify, Program and Verify
  Systems in Rewriting Logic}}.
\newblock {\sl \bibinfo{series}{Lecture Notes of Computer Science}}
  \bibinfo{volume}{4350}, \bibinfo{publisher}{Springer-Verlag},
  \doi{10.1007/978-3-540-71999-1}.

\bibitemdeclare{inproceedings}{Coleman&12a}
\bibitem{Coleman&12a}
\bibinfo{author}{Joey~W. \surnamestart Coleman\surnameend},
  \bibinfo{author}{Anders~Kaels \surnamestart Malmos\surnameend},
  \bibinfo{author}{Peter~Gorm \surnamestart Larsen\surnameend},
  \bibinfo{author}{Jan \surnamestart Peleska\surnameend},
  \bibinfo{author}{Ralph \surnamestart Hains\surnameend}, \bibinfo{author}{Zoe
  \surnamestart Andrews\surnameend}, \bibinfo{author}{Richard \surnamestart
  Payne\surnameend}, \bibinfo{author}{Simon \surnamestart Foster\surnameend},
  \bibinfo{author}{Alvaro \surnamestart Miyazawa\surnameend},
  \bibinfo{author}{Cristiano \surnamestart Bertolini\surnameend} \&
  \bibinfo{author}{Andr{\'{e}} \surnamestart Didier\surnameend}
  (\bibinfo{year}{2012}): \emph{\bibinfo{title}{{COMPASS Tool Vision for a
  System of Systems Collaborative Development Environment}}}.
\newblock In: {\sl \bibinfo{booktitle}{Proceedings of the 7th International
  Conference on System of System Engineering, IEEE SoSE 2012}}, pp.
  \bibinfo{pages}{451--456}, \doi{10.1109/SYSoSE.2012.6384150}.

\bibitemdeclare{inproceedings}{Coleman&12b}
\bibitem{Coleman&12b}
\bibinfo{author}{Joey~W. \surnamestart Coleman\surnameend},
  \bibinfo{author}{Anders~Kaels \surnamestart Malmos\surnameend},
  \bibinfo{author}{Claus~Ballegaard \surnamestart Nielsen\surnameend} \&
  \bibinfo{author}{Peter~Gorm \surnamestart Larsen\surnameend}
  (\bibinfo{year}{2012}): \emph{\bibinfo{title}{{Evolution of the Overture Tool
  Platform}}}.
\newblock In: {\sl \bibinfo{booktitle}{Proceedings of the 10th Overture
  Workshop 2012}}, \bibinfo{series}{School of Computing Science, Newcastle
  University}.

\bibitemdeclare{misc}{20sim}
\bibitem{20sim}
\bibinfo{author}{\surnamestart {Controllab products}\surnameend}
  (\bibinfo{year}{{2013}}): \emph{\bibinfo{title}{{http://www.20sim.com/}}}.
\newblock \bibinfo{note}{{20-Sim official website}}.

\bibitemdeclare{inproceedings}{Couto&13a}
\bibitem{Couto&13a}
\bibinfo{author}{Lu{\'i}s~Diogo \surnamestart Couto\surnameend} \&
  \bibinfo{author}{Richard \surnamestart Payne\surnameend}
  (\bibinfo{year}{2013}): \emph{\bibinfo{title}{{The COMPASS Proof Obligation
  Generator: A test case of Overture Extensibility}}}.
\newblock In: {\sl \bibinfo{booktitle}{Proceedings of the 11th Overture
  Workshop}}.

\bibitemdeclare{inproceedings}{Couto&15b}
\bibitem{Couto&15b}
\bibinfo{author}{Lu{\'i}s~Diogo \surnamestart Couto\surnameend} \&
  \bibinfo{author}{Peter W.~V. \surnamestart Tran-J{\o}rgensen\surnameend}
  (\bibinfo{year}{2015}): \emph{\bibinfo{title}{{Extending the Overture code
  generator towards Isabelle syntax}}}.
\newblock In: {\sl \bibinfo{booktitle}{13th Overture Workshop}},
  \bibinfo{address}{Oslo, Norway}.

\bibitemdeclare{inproceedings}{Couto&15a}
\bibitem{Couto&15a}
\bibinfo{author}{Lu{\'i}s~Diogo \surnamestart Couto\surnameend},
  \bibinfo{author}{Peter W.~V. \surnamestart Tran-J{\o}rgensen\surnameend},
  \bibinfo{author}{Joey~W. \surnamestart Coleman\surnameend} \&
  \bibinfo{author}{Kenneth \surnamestart Lausdahl\surnameend}
  (\bibinfo{year}{2015}): \emph{\bibinfo{title}{{Migrating to an Extensible
  Architecture for Abstract Syntax Trees}}}.
\newblock In: {\sl \bibinfo{booktitle}{12th Working IEEE / IFIP Conference on
  Software Architecture}}.

\bibitemdeclare{misc}{website:dltk}
\bibitem{website:dltk}
\bibinfo{author}{\surnamestart Eclipse\surnameend} (\bibinfo{year}{2015}):
  \emph{\bibinfo{title}{{Dynamic Languages Toolkit}}}.
\newblock \urlprefix\url{http://eclipse.org/dltk/}.

\bibitemdeclare{misc}{EventBWs15}
\bibitem{EventBWs15}
 (\bibinfo{year}{2015}): \emph{\bibinfo{title}{{Event-B and the Rodin
  Platform}}}.
\newblock \bibinfo{howpublished}{\url{http://www.event-b.org}}.

\bibitemdeclare{book}{Fitzgerald&09}
\bibitem{Fitzgerald&09}
\bibinfo{author}{John \surnamestart Fitzgerald\surnameend} \&
  \bibinfo{author}{Peter~Gorm \surnamestart Larsen\surnameend}
  (\bibinfo{year}{2009}): \emph{\bibinfo{title}{{Modelling Systems -- Practical
  Tools and Techniques in Software Development}}}, \bibinfo{edition}{{Second}}
  edition.
\newblock \bibinfo{publisher}{Cambridge University Press},
  \bibinfo{address}{The Edinburgh Building, Cambridge CB2 2RU, UK},
  \doi{10.1017/CBO9780511626975}.
\newblock \bibinfo{note}{{ISBN 0-521-62348-0}}.

\bibitemdeclare{book}{Fitzgerald&05}
\bibitem{Fitzgerald&05}
\bibinfo{author}{John \surnamestart Fitzgerald\surnameend},
  \bibinfo{author}{Peter~Gorm \surnamestart Larsen\surnameend},
  \bibinfo{author}{Paul \surnamestart Mukherjee\surnameend},
  \bibinfo{author}{Nico \surnamestart Plat\surnameend} \&
  \bibinfo{author}{Marcel \surnamestart Verhoef\surnameend}
  (\bibinfo{year}{2005}): \emph{\bibinfo{title}{{Validated Designs for
  Object--oriented Systems}}}.
\newblock \bibinfo{publisher}{Springer, New York}, \doi{10.1007/b138800}.
\newblock \urlprefix\url{http://overturetool.org/publications/books/vdoos/}.

\bibitemdeclare{article}{Fitzgerald&08a}
\bibitem{Fitzgerald&08a}
\bibinfo{author}{John \surnamestart Fitzgerald\surnameend},
  \bibinfo{author}{Peter~Gorm \surnamestart Larsen\surnameend} \&
  \bibinfo{author}{Shin \surnamestart Sahara\surnameend}
  (\bibinfo{year}{2008}): \emph{\bibinfo{title}{{VDMTools: Advances in Support
  for Formal Modeling in VDM}}}.
\newblock {\sl \bibinfo{journal}{ACM Sigplan Notices}}
  \bibinfo{volume}{43}(\bibinfo{number}{2}), pp. \bibinfo{pages}{3--11},
  \doi{10.1145/1361213.1361214}.

\bibitemdeclare{book}{Fitzgerald&14c}
\bibitem{Fitzgerald&14c}
\bibinfo{editor}{John \surnamestart Fitzgerald\surnameend},
  \bibinfo{editor}{Peter~Gorm \surnamestart Larsen\surnameend} \&
  \bibinfo{editor}{Marcel \surnamestart Verhoef\surnameend}, editors
  (\bibinfo{year}{2014}): \emph{\bibinfo{title}{Collaborative Design for
  Embedded Systems -- Co-modelling and Co-simulation}}.
\newblock \bibinfo{publisher}{Springer}, \doi{10.1007/978-3-642-54118-6}.
\newblock
  \urlprefix\url{http://link.springer.com/book/10.1007/978-3-642-54118-6}.

\bibitemdeclare{incollection}{Foster&15}
\bibitem{Foster&15}
\bibinfo{author}{Simon \surnamestart Foster\surnameend}, \bibinfo{author}{Frank
  \surnamestart Zeyda\surnameend} \& \bibinfo{author}{Jim \surnamestart
  Woodcock\surnameend} (\bibinfo{year}{2015}):
  \emph{\bibinfo{title}{{Isabelle/UTP: A mechanised theory engineering
  framework}}}.
\newblock In: {\sl \bibinfo{booktitle}{Unifying Theories of Programming}},
  \bibinfo{publisher}{Springer}, pp. \bibinfo{pages}{21--41},
  \doi{10.1007/978-3-319-14806-9\_2}.

\bibitemdeclare{inproceedings}{Gagnon&98}
\bibitem{Gagnon&98}
\bibinfo{author}{Etienne~M. \surnamestart Gagnon\surnameend} \&
  \bibinfo{author}{Laurie~J. \surnamestart Hendren\surnameend}
  (\bibinfo{year}{1998}): \emph{\bibinfo{title}{{{SableCC}, an Object-Oriented
  Compiler Framework}}}.
\newblock In: {\sl \bibinfo{booktitle}{Proceedings of the Technology of
  Object-Oriented Languages and Systems}}, \bibinfo{series}{TOOLS '98},
  \bibinfo{publisher}{IEEE Computer Society}, \bibinfo{address}{Washington, DC,
  USA}, pp. \bibinfo{pages}{140--154}, \doi{10.1109/TOOLS.1998.711009}.

\bibitemdeclare{book}{DESIGNPAT95}
\bibitem{DESIGNPAT95}
\bibinfo{author}{E.~\surnamestart Gamma\surnameend},
  \bibinfo{author}{R.~\surnamestart Helm\surnameend},
  \bibinfo{author}{R.~\surnamestart Johnson\surnameend} \&
  \bibinfo{author}{R.~\surnamestart Vlissides\surnameend}
  (\bibinfo{year}{1995}): \emph{\bibinfo{title}{Design Patterns. Elements of
  Reusable Object-Oriented Software.}}
\newblock {\sl \bibinfo{series}{Addison-Wesley Professional Computing
  Series}}~, \bibinfo{publisher}{Addison-Wesley Publishing Company}.

\bibitemdeclare{article}{Hoare78}
\bibitem{Hoare78}
\bibinfo{author}{C.A.R \surnamestart Hoare\surnameend} (\bibinfo{year}{1978}):
  \emph{\bibinfo{title}{{Communicating Sequential Processes}}}.
\newblock {\sl \bibinfo{journal}{Communications of the ACM}}
  \bibinfo{volume}{21}(\bibinfo{number}{8}), \doi{10.1145/359576.359585}.

\bibitemdeclare{book}{Hoare&98}
\bibitem{Hoare&98}
\bibinfo{author}{Tony \surnamestart Hoare\surnameend} \&
  \bibinfo{author}{He~\surnamestart Jifeng\surnameend} (\bibinfo{year}{1998}):
  \emph{\bibinfo{title}{Unifying Theories of Programming}}.
\newblock \bibinfo{publisher}{Prentice Hall}, \doi{10.1007/11768173}.

\bibitemdeclare{misc}{website:isabelle-eclipse}
\bibitem{website:isabelle-eclipse}
\bibinfo{author}{\surnamestart Isabelle/Eclipse\surnameend}
  (\bibinfo{year}{2015}): \emph{\bibinfo{title}{Isabelle/Eclipse}}.
\newblock \urlprefix\url{http://andriusvelykis.github.io/isabelle-eclipse/}.

\bibitemdeclare{incollection}{Jackson&09doi}
\bibitem{Jackson&09doi}
\bibinfo{author}{Ethan~K. \surnamestart Jackson\surnameend},
  \bibinfo{author}{Dirk \surnamestart Seifert\surnameend},
  \bibinfo{author}{Markus \surnamestart Dahlweid\surnameend},
  \bibinfo{author}{Thomas \surnamestart Santen\surnameend},
  \bibinfo{author}{Nikolaj \surnamestart Bj\o{}rner\surnameend} \&
  \bibinfo{author}{Wolfram \surnamestart Schulte\surnameend}
  (\bibinfo{year}{2009}): \emph{\bibinfo{title}{Specifying and Composing
  Non-functional Requirements in Model-Based Development}}.
\newblock In \bibinfo{editor}{Alexandre \surnamestart Bergel\surnameend} \&
  \bibinfo{editor}{Johan \surnamestart Fabry\surnameend}, editors: {\sl
  \bibinfo{booktitle}{Software Composition}}, {\sl \bibinfo{series}{Lecture
  Notes in Computer Science}} \bibinfo{volume}{5634},
  \bibinfo{publisher}{Springer Berlin Heidelberg}, pp. \bibinfo{pages}{72--89},
  \doi{10.1007/978-3-642-02655-3\_7}.

\bibitemdeclare{inproceedings}{Jones99}
\bibitem{Jones99}
\bibinfo{author}{Cliff~B.\ \surnamestart Jones\surnameend}
  (\bibinfo{year}{1999}): \emph{\bibinfo{title}{{Scientific Decisions which
  Characterize VDM}}}.
\newblock In \bibinfo{editor}{J.M.\ \surnamestart Wing\surnameend},
  \bibinfo{editor}{J.C.P.\ \surnamestart Woodcock\surnameend} \&
  \bibinfo{editor}{J.\ \surnamestart Davies\surnameend}, editors: {\sl
  \bibinfo{booktitle}{FM'99 - Formal Methods}},
  \bibinfo{publisher}{Springer-Verlag}, pp. \bibinfo{pages}{28--47},
  \doi{10.1007/3-540-48119-2\_2}.
\newblock \bibinfo{note}{Lecture Notes in Computer Science 1708}.

\bibitemdeclare{inproceedings}{Jorgensen&14a}
\bibitem{Jorgensen&14a}
\bibinfo{author}{Peter~W.V. \surnamestart J\o{}rgensen\surnameend},
  \bibinfo{author}{Lu{\'i}s~D. \surnamestart Couto\surnameend} \&
  \bibinfo{author}{Morten \surnamestart Larsen\surnameend}
  (\bibinfo{year}{2014}): \emph{\bibinfo{title}{{A Code Generation Platform for
  VDM}}}.
\newblock In: {\sl \bibinfo{booktitle}{The Overture 2014 workshop}}.

\bibitemdeclare{book}{TLAplus}
\bibitem{TLAplus}
\bibinfo{author}{Leslie \surnamestart Lamport\surnameend}
  (\bibinfo{year}{2002}): \emph{\bibinfo{title}{Specifying Systems, The {TLA}+
  Language and Tools for Hardware and Software Engineers}}.
\newblock \bibinfo{publisher}{Addison-Wesley}.
\newblock
  \urlprefix\url{http://research.microsoft.com/users/lamport/tla/book.html}.

\bibitemdeclare{misc}{ISOVDM96short}
\bibitem{ISOVDM96short}
\bibinfo{author}{P.~G. \surnamestart Larsen\surnameend}, \bibinfo{author}{B.~S.
  \surnamestart Hansen\surnameend} et~al. (\bibinfo{year}{1996}):
  \emph{\bibinfo{title}{{Information technology -- Programming languages, their
  environments and system software interfaces -- Vienna Development Method --
  Specification Language -- Part 1: Base language}}}.
\newblock \bibinfo{note}{{International Standard ISO/IEC 13817-1}}.

\bibitemdeclare{article}{Larsen&10adoi}
\bibitem{Larsen&10adoi}
\bibinfo{author}{Peter~Gorm \surnamestart Larsen\surnameend},
  \bibinfo{author}{Nick \surnamestart Battle\surnameend},
  \bibinfo{author}{Miguel \surnamestart Ferreira\surnameend},
  \bibinfo{author}{John \surnamestart Fitzgerald\surnameend},
  \bibinfo{author}{Kenneth \surnamestart Lausdahl\surnameend} \&
  \bibinfo{author}{Marcel \surnamestart Verhoef\surnameend}
  (\bibinfo{year}{2010}): \emph{\bibinfo{title}{{The Overture Initiative --
  Integrating Tools for VDM}}}.
\newblock {\sl \bibinfo{journal}{SIGSOFT Softw. Eng. Notes}}
  \bibinfo{volume}{35}(\bibinfo{number}{1}), pp. \bibinfo{pages}{1--6},
  \doi{10.1145/1668862.1668864}.

\bibitemdeclare{inproceedings}{Larsen&10cdoi}
\bibitem{Larsen&10cdoi}
\bibinfo{author}{Peter~Gorm \surnamestart Larsen\surnameend},
  \bibinfo{author}{Kenneth \surnamestart Lausdahl\surnameend} \&
  \bibinfo{author}{Nick \surnamestart Battle\surnameend}
  (\bibinfo{year}{2010}): \emph{\bibinfo{title}{{Combinatorial Testing for
  VDM}}}.
\newblock In: {\sl \bibinfo{booktitle}{Proceedings of the 2010 8th IEEE
  International Conference on Software Engineering and Formal Methods}},
  \bibinfo{series}{SEFM '10}, \bibinfo{publisher}{IEEE Computer Society},
  \bibinfo{address}{Washington, DC, USA}, pp. \bibinfo{pages}{278--285},
  \doi{10.1109/SEFM.2010.32}.
\newblock \bibinfo{note}{{ISBN 978-0-7695-4153-2}}.

\bibitemdeclare{inproceedings}{Lausdahl&14a}
\bibitem{Lausdahl&14a}
\bibinfo{author}{Kenneth \surnamestart Lausdahl\surnameend},
  \bibinfo{author}{Hiroshi \surnamestart Ishikawa\surnameend} \&
  \bibinfo{author}{Peter~Gorm \surnamestart Larsen\surnameend}
  (\bibinfo{year}{2015}): \emph{\bibinfo{title}{{Interpreting Implicit VDM
  Specifications using ProB}}}.
\newblock In: {\sl \bibinfo{booktitle}{Proceedings of the 12th Overture
  Workshop}}, {\sl \bibinfo{series}{Technical Report Series}}
  \bibinfo{volume}{CS-TR-1446}, \bibinfo{organization}{Computing Science,
  Newcastle University}, pp. \bibinfo{pages}{1--15}.
\newblock
  \urlprefix\url{http://www.cs.ncl.ac.uk/publications/trs/papers/1446.pdf}.

\bibitemdeclare{inproceedings}{Lausdahl&11}
\bibitem{Lausdahl&11}
\bibinfo{author}{Kenneth \surnamestart Lausdahl\surnameend},
  \bibinfo{author}{Peter~Gorm \surnamestart Larsen\surnameend} \&
  \bibinfo{author}{Nick \surnamestart Battle\surnameend}
  (\bibinfo{year}{2011}): \emph{\bibinfo{title}{{A Deterministic Interpreter
  Simulating A Distributed real time system using VDM}}}.
\newblock In \bibinfo{editor}{Shengchao \surnamestart Qin\surnameend} \&
  \bibinfo{editor}{Zongyan \surnamestart Qiu\surnameend}, editors: {\sl
  \bibinfo{booktitle}{Proceedings of the 13th international conference on
  Formal methods and software engineering}}, {\sl \bibinfo{series}{Lecture
  Notes in Computer Science}} \bibinfo{volume}{6991},
  \bibinfo{publisher}{Springer-Verlag}, \bibinfo{address}{Berlin, Heidelberg},
  pp. \bibinfo{pages}{179--194}, \doi{10.1007/978-3-642-24559-6\_14}.
\newblock \urlprefix\url{http://dl.acm.org/citation.cfm?id=2075089.2075107}.
\newblock \bibinfo{note}{{ISBN 978-3-642-24558-9}}.

\bibitemdeclare{incollection}{leuschel&03}
\bibitem{leuschel&03}
\bibinfo{author}{Michael \surnamestart Leuschel\surnameend} \&
  \bibinfo{author}{Michael \surnamestart Butler\surnameend}
  (\bibinfo{year}{2003}): \emph{\bibinfo{title}{ProB: A model checker for B}}.
\newblock In: {\sl \bibinfo{booktitle}{FME 2003: Formal Methods}},
  \bibinfo{publisher}{Springer}, pp. \bibinfo{pages}{855--874},
  \doi{10.1007/978-3-540-45236-2\_46}.

\bibitemdeclare{incollection}{leuschel&05}
\bibitem{leuschel&05}
\bibinfo{author}{Michael \surnamestart Leuschel\surnameend} \&
  \bibinfo{author}{Michael \surnamestart Butler\surnameend}
  (\bibinfo{year}{2005}): \emph{\bibinfo{title}{Automatic refinement checking
  for B}}.
\newblock In: {\sl \bibinfo{booktitle}{Formal Methods and Software
  Engineering}}, \bibinfo{publisher}{Springer}, pp. \bibinfo{pages}{345--359},
  \doi{10.1007/11576280\_24}.

\bibitemdeclare{inproceedings}{Mukherjee&00}
\bibitem{Mukherjee&00}
\bibinfo{author}{Paul \surnamestart Mukherjee\surnameend},
  \bibinfo{author}{Fabien \surnamestart Bousquet\surnameend},
  \bibinfo{author}{J\'{e}r\^{o}me \surnamestart Delabre\surnameend},
  \bibinfo{author}{Stephen \surnamestart Paynter\surnameend} \&
  \bibinfo{author}{Peter~Gorm \surnamestart Larsen\surnameend}
  (\bibinfo{year}{2000}): \emph{\bibinfo{title}{{Exploring Timing Properties
  Using VDM++ on an Industrial Application}}}.
\newblock In \bibinfo{editor}{J.C. \surnamestart Bicarregui\surnameend} \&
  \bibinfo{editor}{J.S. \surnamestart Fitzgerald\surnameend}, editors: {\sl
  \bibinfo{booktitle}{Proceedings of the Second VDM Workshop}}.
\newblock \bibinfo{note}{{Available at www.vdmportal.org}}.

\bibitemdeclare{book}{nipkow02}
\bibitem{nipkow02}
\bibinfo{author}{Tobias \surnamestart Nipkow\surnameend},
  \bibinfo{author}{Lawrence~C \surnamestart Paulson\surnameend} \&
  \bibinfo{author}{Markus \surnamestart Wenzel\surnameend}
  (\bibinfo{year}{2002}): \emph{\bibinfo{title}{{Isabelle/HOL: a proof
  assistant for higher-order logic}}}.
\newblock \bibinfo{volume}{2283}, \bibinfo{publisher}{Springer Science \&
  Business Media}, \doi{10.1007/3-540-45949-9}.

\bibitemdeclare{book}{Parr07}
\bibitem{Parr07}
\bibinfo{author}{Terence \surnamestart Parr\surnameend} (\bibinfo{year}{2007}):
  \emph{\bibinfo{title}{The Definitive ANTLR Reference: Building
  Domain-Specific Languages}}.
\newblock \bibinfo{publisher}{Pragmatic Bookshelf}.

\bibitemdeclare{inproceedings}{Paulson10}
\bibitem{Paulson10}
\bibinfo{author}{Lawrence~C. \surnamestart Paulson\surnameend}
  (\bibinfo{year}{2010}): \emph{\bibinfo{title}{{Three Years of Experience with
  Sledgehammer, a Practical Link between Automatic and Interactive Theorem
  Provers}}}.
\newblock In \bibinfo{editor}{Renate~A. \surnamestart Schmidt\surnameend},
  \bibinfo{editor}{Stephan \surnamestart Schulz\surnameend} \&
  \bibinfo{editor}{Boris \surnamestart Konev\surnameend}, editors: {\sl
  \bibinfo{booktitle}{Proceedings of the 2nd Workshop on Practical Aspects of
  Automated Reasoning, PAAR-2010, Edinburgh, Scotland, UK, July 14, 2010}},
  {\sl \bibinfo{series}{EPiC Series}}~\bibinfo{volume}{9}, pp.
  \bibinfo{pages}{1--10}.

\bibitemdeclare{inproceedings}{peleska&11}
\bibitem{peleska&11}
\bibinfo{author}{Jan \surnamestart Peleska\surnameend}, \bibinfo{author}{Elena
  \surnamestart Vorobev\surnameend} \& \bibinfo{author}{Florian \surnamestart
  Lapschies\surnameend} (\bibinfo{year}{2011}):
  \emph{\bibinfo{title}{{Automated Test Case Generation with SMT-Solving and
  Abstract Interpretation}}}.
\newblock In \bibinfo{editor}{Mihaela \surnamestart Bobaru\surnameend},
  \bibinfo{editor}{Klaus \surnamestart Havelund\surnameend},
  \bibinfo{editor}{Gerard~J. \surnamestart Holzmann\surnameend} \&
  \bibinfo{editor}{Rajeev \surnamestart Joshi\surnameend}, editors: {\sl
  \bibinfo{booktitle}{Nasa Formal Methods, Third International Symposium, NFM
  2011}}, \bibinfo{organization}{NASA}, \bibinfo{publisher}{Springer LNCS
  6617}, \bibinfo{address}{Pasadena, CA, USA}, pp. \bibinfo{pages}{298--312},
  \doi{10.1007/978-3-642-20398-5\_22}.

\bibitemdeclare{misc}{TlaWS15}
\bibitem{TlaWS15}
 (\bibinfo{year}{2015}): \emph{\bibinfo{title}{{The TLA Toolbox}}}.
\newblock
  \bibinfo{howpublished}{\url{http://research.microsoft.com/en-us/um/people/lamport/tla/toolbox.html}}.

\bibitemdeclare{phdthesis}{Verhoef09}
\bibitem{Verhoef09}
\bibinfo{author}{Marcel \surnamestart Verhoef\surnameend}
  (\bibinfo{year}{2009}): \emph{\bibinfo{title}{{Modeling and Validating
  Distributed Embedded Real-Time Control Systems}}}.
\newblock Ph.D. thesis, \bibinfo{school}{Radboud University Nijmegen}.

\bibitemdeclare{inproceedings}{Verhoef&06b}
\bibitem{Verhoef&06b}
\bibinfo{author}{Marcel \surnamestart Verhoef\surnameend},
  \bibinfo{author}{Peter~Gorm \surnamestart Larsen\surnameend} \&
  \bibinfo{author}{Jozef \surnamestart Hooman\surnameend}
  (\bibinfo{year}{2006}): \emph{\bibinfo{title}{{Modeling and Validating
  Distributed Embedded Real-Time Systems with VDM++}}}.
\newblock In \bibinfo{editor}{Jayadev \surnamestart Misra\surnameend},
  \bibinfo{editor}{Tobias \surnamestart Nipkow\surnameend} \&
  \bibinfo{editor}{Emil \surnamestart Sekerinski\surnameend}, editors: {\sl
  \bibinfo{booktitle}{FM 2006: Formal Methods}}, \bibinfo{series}{Lecture Notes
  in Computer Science 4085}, \bibinfo{publisher}{Springer-Verlag}, pp.
  \bibinfo{pages}{147--162}, \doi{10.1007/11813040\_11}.

\bibitemdeclare{inproceedings}{Woodcock&12a}
\bibitem{Woodcock&12a}
\bibinfo{author}{J.~\surnamestart Woodcock\surnameend},
  \bibinfo{author}{A.~\surnamestart Cavalcanti\surnameend},
  \bibinfo{author}{J.~\surnamestart Fitzgerald\surnameend},
  \bibinfo{author}{P.~\surnamestart Larsen\surnameend},
  \bibinfo{author}{A.~\surnamestart Miyazawa\surnameend} \&
  \bibinfo{author}{S.~\surnamestart Perry\surnameend} (\bibinfo{year}{2012}):
  \emph{\bibinfo{title}{{Features of CML: a Formal Modelling Language for
  Systems of Systems}}}.
\newblock In: {\sl \bibinfo{booktitle}{Proceedings of the 7th International
  Conference on System of System Engineering}}, \bibinfo{publisher}{IEEE},
  \doi{10.1109/SYSoSE.2012.6384144}.

\bibitemdeclare{misc}{website:xtest}
\bibitem{website:xtest}
\bibinfo{author}{\surnamestart Xtext\surnameend} (\bibinfo{year}{2015}):
  \emph{\bibinfo{title}{Xtext}}.
\newblock \urlprefix\url{https://eclipse.org/Xtext/}.

\end{thebibliography}

\end{document}